\documentclass[twocolumn,twoside,slac_two]{revtex4}
\usepackage{graphicx}
\usepackage{fancyhdr}
\begin{document}

\title{Fermi-LAT Upper Limit for NGC 4151 and its Implications for Physics of Hot Accretion Flow}

\author{R. Wojaczy\'nski, A. Nied\'zwiecki}
\affiliation{Department of Astrophysics, University of \L \'od\'z, Pomorska 149/153, 
90-236 \L \'od\'z, Poland}
\author{F.-G. Xie}
\affiliation{Key Laboratory for Research in Galaxies and Cosmology, Shanghai
  Astronomical Observatory, Chinese Academy of Sciences,
80 Nandan Road, Shanghai 200030, China}

\begin{abstract}
We present preliminary results of our analysis of the {\it Fermi}-LAT data from the
direction of NGC 4151. We find a new $\gamma$-ray source with a statistical significance $\sigma > 5$, shifted by 
$0.5^\circ$ from the position of NGC 4151. Apparently, the source was bright only during a 1.5-year period between
December 2011 and June 2013 and it strongly contaminated the signal from NGC 4151. Therefore, we 
neglect this period in our analysis. We find two additional, persistent $\gamma$-ray sources  with high $\sigma$, shifted from   NGC 4151 by $\sim 1.5^\circ$ and $5^\circ$,
whose presence has been recently confirmed in the Third Fermi Catalog. 
After subtracting the above sources, we still see a weak residual, with $\sigma \lesssim 3$, at the position
of NGC 4151.
We derive an upper limit (UL) for the $\gamma$-ray flux from NGC 4151 and we compare it with
predictions of the ADAF model which can explain the X-ray observations of this
object. We find that the
{\it Fermi} UL strongly constrains non-thermal acceleration processes in hot
flows as well as the values of some crucial parameters.
Here we present the comparison with the hot flow models in which heating of electrons is dominated
by Coulomb interactions with hot protons. In such a version of the model, the $\gamma$-ray UL, combined with 
the X-ray data,  constrains the energy content in the non-thermal component of proton
distribution to at most a few per cent,  rules out a weak (sub-equipartition) magnetic field and favors a rapid rotation of the supermassive black hole.

\end{abstract}

\maketitle

\thispagestyle{fancy}

\section{INTRODUCTION}

Low-luminosity AGNs, with the luminosities below $\sim 0.01 L_{\rm Edd}$, are likely to be powered by optically thin, hot accretion flows (a.k.a.\ ADAFs, see e.g.\ [12]). The two-temperature structure is a key property of ADAFs, as such flows are supported by the proton pressure. In their innermost parts, the hot protons have energies above the threshold for pion production. As estimated e.g.\ in [6,8],
the decay of pions leads to substantial fluxes of $\gamma$-rays, which may be probed in nearby AGNs at the current sensitivity of {\it Fermi}-LAT surveys. 
The  {\it Fermi}-LAT data for radio-quiet AGNs were analyzed in [10] (two years of data) and [2] (three years) and the derived upper limits are already quite stringent compared to expectation. In [11]we revisit the issue of searching the signatures of hadronic emission from hot flows and the related implications for hot-flow models. 
We perform the detailed analysis of nearby, low-luminosity Seyfert galaxies using over 6 years of  {\it Fermi}-LAT data and we compare the results with the model predictions for a complete range of the model parameters. In this contribution we report our preliminary results for one of the best-studied AGNs, NGC 4151.

NGC 4151 is one the X-ray brightest AGNs, with the bolometric $L \sim 0.01 L_{\rm Edd}$,  showing  no signatures of a relativistically distorted reflection component (constraints on the width of Fe K$\alpha$ line imply the lack of an optically thick material within at least the innermost $\sim 100 R_{\rm g}$) as well as showing a spectral similarity to black hole binaries in their hard states (also most likely powered by hot flows), see e.g.\ [4]. All these properties make it a relevant objects for testing the hot flow scenario. Below we use the black hole mass $M=3.8 \times 10^7M_\odot$ from the stellar dynamical mass measurement [9].

\begin{figure*}[t]
\centerline{\includegraphics[width=75mm]{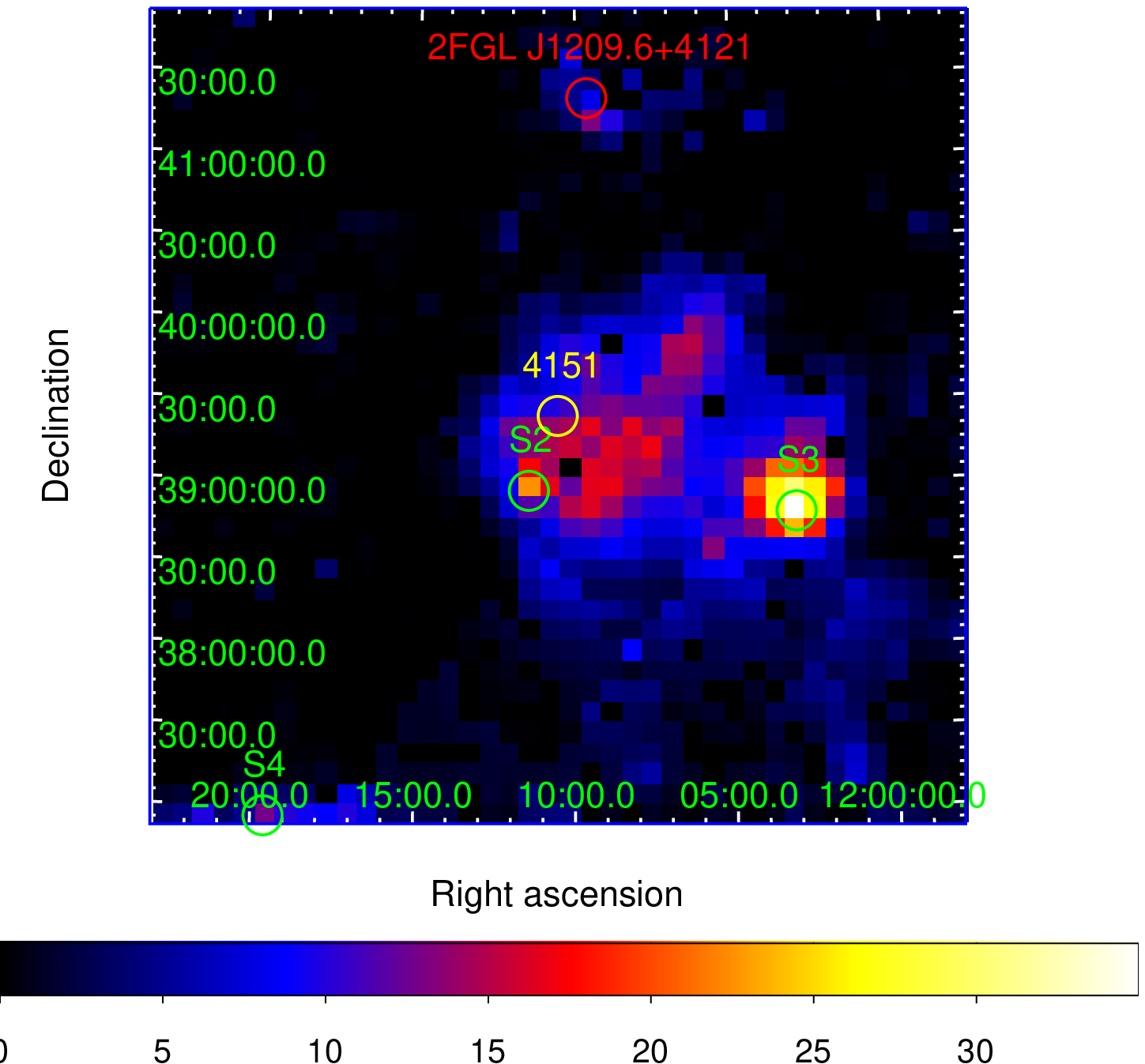} \hspace{0.5cm} \includegraphics[width=75mm]{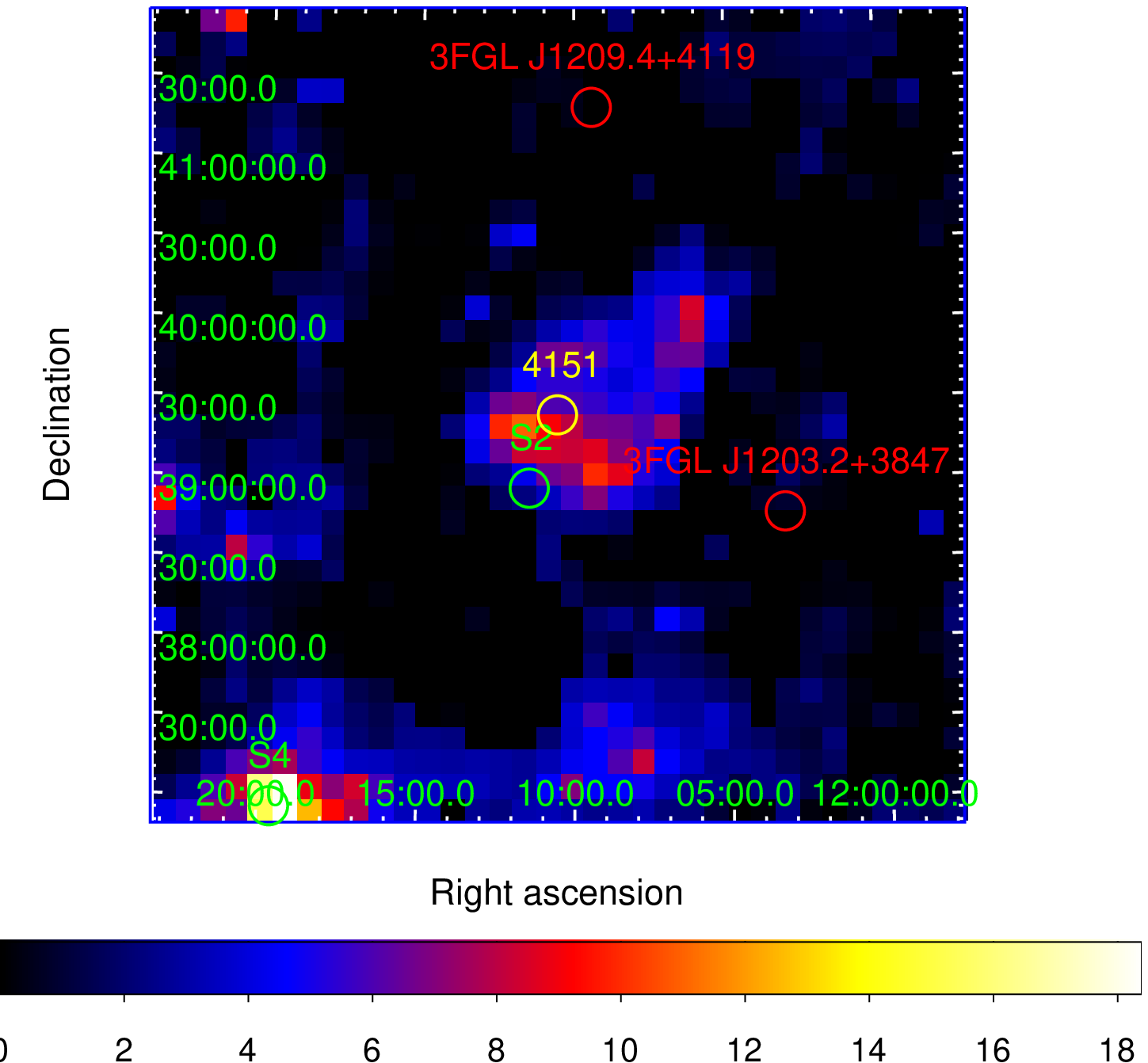}}
\caption{Test-statistic maps of the $5^\circ \times 5^\circ$ region around the
position of NGC 4151 (yellow circle) with the pixel size of $0.125^\circ$ (left) and  $0.15^\circ$ (right). The left map was generated from the data from the total 6.4 years and only the sources from  2FGL were subtracted from this map (the only 2FGL source in this region of the sky is marked by the red circle); the green circles indicate new sources revealed in our analysis. The right map was generated from the data taken before December 2011 and after June 2013 and  the sources from 3FGL (red circles) were subtracted from this map.} 
\label{fig1}
\end{figure*}

\section{LAT DATA ANALYSIS}

We analyzed the data from the direction of NGC 4151, comprising 6.4 years of Fermi-LAT observations carried out between 2008 August 4 and 2015 January 10. Events we selected from a region with the radius of $10^\circ$ centered on the position of NGC 4151. We performed the unbinned likelihood analysis using the v9r33p0 {\it Fermi} Science Tools with CALDB instrument response functions.  We used the standard templates for the Galactic (\texttt{gll\_iem\_v05\_rev1.fits}) and the isotropic (\texttt{iso\_source\_v05\_rev1.txt}) backgrounds. 
In our initial model of the region we took into account only the sources from the Second Fermi Catalog (2FGL) [7], i.e.\ our model included the same sources as those used in [2]. 2FGL J1209.6+4121 (marked by the red circle in Fig.\ 1a) is the 2FGL source closest to NGC 4151,  with the distance of $\simeq 2^\circ$.

\begin{table}[t]
\begin{center}
\caption{New sources introduced in the model of the region around NGC 4151, not reported
in 2FGL. (1) Source (see Fig.\ 1), (2) 3FGL name, (3) \texttt{gtlike} TS values, (4) and (5) \texttt{gtfindsrc} coordinates.
Results for S2 were obtained using the data taken between December 2011 and June 2013; for the remaining sources the total data set for 6.4 years was used.
}
\begin{tabular}{|c|c|c|c|c|c|}
\hline 
(1) & (2) 3FGL name & (3) TS & (4) RA & (5) DEC \\
\hline 
S1 & 3FGL J1220.2+3434 & 191 & 185.06 & 34.57 \\
S2 & not reported & 30 &  182.86 & 38.95 \\
S3 & 3FGL J1203.2+3847 & 32 & 180.81 & 38.79 \\
S4 & not reported & 22 & 184.90 & 36.93 \\
\hline
\end{tabular}
\label{tab:s}
\end{center}
\end{table}

Fig.\ 1a shows the TS map of the region, built after subtracting the 2FGL sources.
The map reveals residual structures indicating the presence of additional point-like sources, marked by the green circles.
For each of these objects we use  the  \texttt{gtlike} and \texttt{gtfindsrc} tools to find its significance,  best-fit position and 
spectral parameters; the results are given in Table \ref{tab:s}. 
Sources S1 (beyond the map in Fig.\ 1) and S3 have been recently reported in 3FGL [1], with parameters very similar to these estimated in our analysis. 
S2 is not reported in 3FGL, however, this source is critical for the analysis of a signal from NGC 4151, as its distance of $\sim 0.5^\circ$ is comparable (or smaller below $\sim 1$ GeV)
to the LAT point spread function. Therefore, we check properties of this source in more details.

 By using \texttt{gtsrcprob} we find that the position of S2 is determined mostly by 4 photons with energies between 10 and 20 GeV
which arrived from the same direction (within 10 arcmin) between December 2011 and June 2013. At lower energies, neglecting the four events with $E > 10$ GeV, we also see the signature of increased activity of S2 during that period, in the form of an extended residual covering the nominal positions of NGC 4151 and S2.

In the TS map built for the data neglecting the above period (see Fig.\ 1b) we do not see a strong signal at the S2 position, we therefore conclude that S2 strongly dominated the emission from the region around NGC 4151 only during the 1.5 year out of the total 6.4 considered years. 
Note that S2 is not reported  in 3FGL which includes sources detected with $TS > 25$  using the data taken during the four years up to 2012 July  (i.e.\ covering only $\sim 30\%$ of the time of the increased activity of S2).

For our further analysis we neglect the data taken during the 1.5 year when S2 was bright. We subtract the sources reported in  3FGL and we get the TS map shown Fig.\ 1b. The map shows a weak residual, which can be fully compensated for by adding the source at the nominal position of NGC 4151 and \texttt{gtlike}  gives $TS \simeq 8$ for such a source.
At its very low statistical significance, 
it is not possible to assess whether it represents a background fluctuation or an actual emission
from the studied object; it may also contain some contribution from emission of S2 in its lower luminosity states.

\begin{figure*}[t]
\centering
\includegraphics[width=125mm]{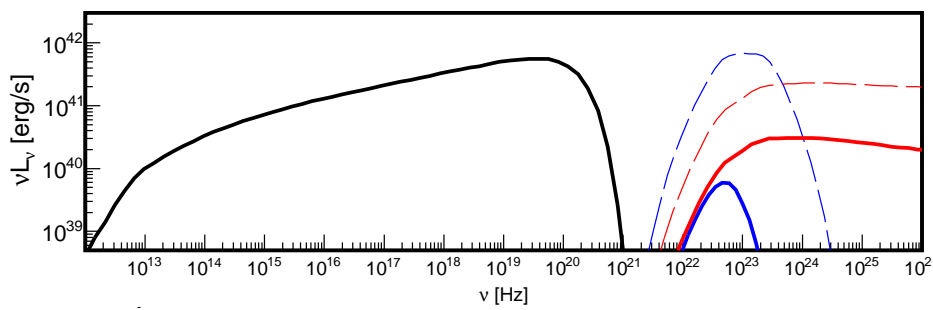}
\caption{Spectra received by a distant observer from a hot flow around the black hole with $M=4 \times 10^7M_\odot$ for $a=0.95$, $\beta=9$, $\delta=10^{-3}$ and $\dot m = 0.3$. The black solid line shows the component formed by the cooling of electrons (mostly thermal Comptonization). The blue and red solid lines show the component formed by $\pi^0$-decay for thermal and power-law (with $s=2.1$) distribution of protons, respectively; the thin dashed lines show the corresponding spectra of the hadronic component in the rest frame of the flow (the spectra shown by solid lines are affected by the GR transfer and  $\gamma \gamma$ absorption).} 
\label{fig2}
\end{figure*}

\begin{table}[t]
\begin{center}
\caption{95\% C.L.\ upper limits for the photon flux (3) and $\gamma$-ray luminosity (4) for the energy ranges given in column (1) and the assumed value of $\gamma$ given in column(2).}
\begin{tabular}{|c|c|c|c|}
\hline 
(1) Energy & (2) Photon & (3) UL: $F$ & (4) UL: $L_{\gamma}$ \\
range [GeV]& index & [phot/cm$^2$/s] & [erg/s] \\
\hline 
0.3 - 1 & 4 & $6 \times 10^{-10}$ & $6.5 \times 10^{39}$ \\
1 - 10 & 2.1  & $1.5 \times 10^{-10}$ & $3.1 \times 10^{39}$ \\
1 - 10 & 2.7 & $1.3 \times 10^{-10}$ & $1.9 \times 10^{39}$ \\
\hline
\end{tabular}
\label{tab:ul}
\end{center}
\end{table}

We then derive the 95\% confidence level upper limit (UL) for the integrated photon flux from NGC 4151 neglecting the data between December 2011 and June 2013.
The pion decay spectra can be approximated by a simple power-law only in limited energy ranges (see Fig.\ 2 below), therefore, we assume relevant values of the photon index, $\Gamma$, and find the UL in the 0.3--1 GeV range to compare with the $\pi^0$-decay spectra for the thermal distribution of protons, and in the 1--10 GeV to compare with the model assuming a power-law distribution of protons. The results are given in Table \ref{tab:ul}.

\section{GAMMA-RAY EMISSION FROM HOT FLOWS AND COMPARISON WITH NGC 4151}

In [5,6] we present a precise model for emission from hot flows, taking into account the relevant leptonic and hadronic processes, and using 
(1) a fully general relativistic (GR) description of both the radiative and hydrodynamic processes; (2) an exact, Monte Carlo computation of global Comptonization; (3) seed photons input from nonthermal synchrotron emission of $\pi^\pm$-decay electrons; (4) an exact computation of the absorption of $\gamma$-ray photons in the radiation field of the flow.
The model is parametrized by: the black hole mass, $M$, the dimensionless accretion rate, $\dot m = \dot M c^2/ L_{\rm Edd}$, the ratio of gas to magnetic pressure, $\beta$,
the fraction of the dissipated energy which heats directly electrons through MHD processes, $\delta$, and the spin parameter, $a = J / (c R_{\rm g} M)$, where $J$ is the  black hole angular momentum and $R_{\rm g}=GM/c^2$. We take into account models with 
thermal and with power-law distrubutions of protons; in the latter the power-law index, $s$, is also a free parameter.

Here we focus on models with small $\delta$, i.e.\ with electrons heated by Coulomb interactions.
We briefly summarize  properties crucial for our final conclusions; Fig.\ \ref{fig2} shows example spectra of radiation produced in a hot flow by thermal Comptonization and by $\pi^0$ decay.

\noindent
(i) The nonthermal synchrotron radiation from $\pi^\pm$-decay electrons gives the dominating input of seed photons for Comptonization and it allows to reconcile the hot-flow model with the AGN X-ray data. It also provides 
an attractive explanation of spectral differences between AGNs
and black-hole transients within the same physical model, see [5].

\noindent
(ii) For bolometric  $L \sim (0.001-0.01) L_{\rm Edd}$, the size of the $\gamma$-ray photosphere (inside which the flow is  opaque to $\gamma$-rays) equals several $R_{\rm g}$. As a result, for models assuming thermal protons, the $\gamma$-ray flux detected by a distant observer is reduced by several orders of magnitude, because the $\gamma$-rays are produced mostly inside the photosphere. We note that [8] assessed comparable X-ray and $\gamma$-ray fluxes from flows surrounding rapidly rotating black holes. However, they neglected the GR transfer and $\gamma \gamma$ absorption;
taking into account these effects we get, for thermal protons, the $\gamma$-ray fluxes smaller by $\sim 3$ (large $\beta$) to $\sim 5$ (small $\beta$) orders of magnitude than the X-ray flux.

\noindent
(iii) In models assuming a thermal distribution of protons, the $\gamma$-ray flux is extremely sensitive to the value of $\beta$.
In flows with smaller $\beta$ (larger $B$), a larger fraction of the accretion power is used to build up
the magnetic field strength; therefore, the energy heating the particles, and hence the proton temperature, is smaller.
As for thermal protons the $\gamma$-ray luminosity, $L_{\gamma}$, is extremely sensitive to the proton temperature, the above effect leads to the difference by 2 -- 3 orders of magnitude between $L_{\gamma}$ for a strong (in equipartition with gas) and weak magnetic field.

\noindent
(iv)  The proton temperature increases with $a$ and hence the $\gamma$-ray emissivities strongly depends on $a$ for thermal protons. However, the largest difference between the emissivities  occurs within the photosphere and, therefore, the dependence of the observed $L_{\gamma}$ on $a$ is reduced by $\gamma \gamma$ absorption.  
On the other hand, a sufficiently strong input of seed photons from the emission of $\pi^{\pm}$-decay  electrons requires either a rapid rotation of the black hole or a significant content of nonthermal protons.

\begin{figure}[t]
\centering
\includegraphics[width=70mm]{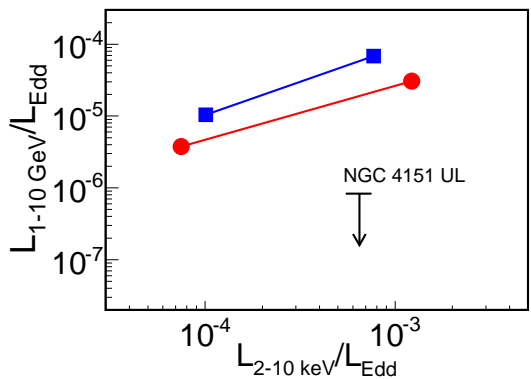}
\caption{The $\gamma$-ray (1--10 GeV) luminosity Eddington ratio as a function of the X-ray (2--10 ken) luminosity Eddington ratio.
The squares and circles show the hot flow model predictions for  the {\it nonthermal proton distribution} with $s=2.1$. The {\it Fermi} UL was obtained for the assumed $\Gamma=2.1$ in the 1--10 GeV range (see Table 2) and the average X-ray luminosity was estimated from the  {\it Swift}-BAT data.
The red circles are for $a=0.998$, $\beta=1$, $\dot m=0.1$ and 0.3; the blue squares are for $a=0.95$, $\beta=9$, $\dot m=0.3$ and 0.8.} 
\label{fig3}
\end{figure}

\begin{figure}[t]
\centering
\includegraphics[width=70mm]{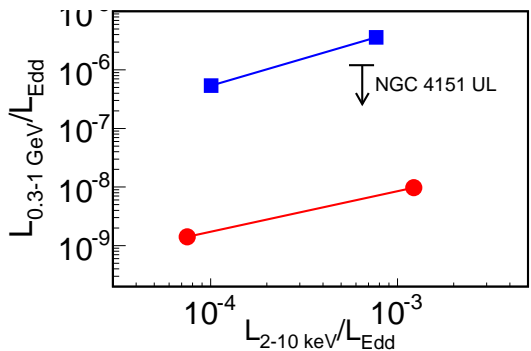}
\caption{The same as in Fig.\ 3 but the model points are for the {\it thermal} distribution of protons and the $\gamma$-ray  Eddington ratio is determined for the 0.3--1 GeV range; $\Gamma=4$ was assumed
for the {\it Fermi} UL.} 
\label{fig4}
\end{figure}

Using the publicly available {\it Swift}-BAT data [3] we find the average X-ray luminosity of NGC 4151 between 2009 and 2014 (we find that it corresponds to the intermediate state as defined in [4]), which then allows us to compare the $L_{\gamma}$ predicted by the model with the {\it Fermi} UL. The results are shown in Figs 3 and 4. For the power-law distribution of protons (Fig.\ 3) the UL is over an order of magnitude lower than the model prediction, which constrains the energy content in the nonthermal component of the proton distribution to at most a few per cent. 

For the thermal distribution of protons (Fig.\ 4), the {\it Fermi} UL is sufficiently low to exlude models with large $\beta$, which, taking into account the above, rules out any version of the model with a weak magnetic field. For an equipartition value of $\beta \sim 1$ the predicted flux is below the UL value.

Then, the {\it Fermi} data favor a strongly magnetized plasma with a weak content of nonthermal protons. For such a case, a high spin value is required for a sufficiently strong flux of seed photons from nonthermal emission of pion-decay electrons.

\section{SUMMARY}

We thoroughly analyzed the $\gamma$-ray data from a region around NGC 4151, which led us to identification of 
new $\gamma$-ray sources. After subtracting their contribution, we get a weak 
residual signal at the position of NGC 4151. At its low statistical significance, $\sigma \lesssim 3$, it is
not possible to assess its nature.

Comparison of the derived upper limits with the model predictions allows to constrain several crucial quantities
which illustrates the potential of {\it Fermi} measurements in probing the properties of flows powering AGNs at low
luminosities.

\bigskip 
\begin{acknowledgments}
This research has been supported in part by the Polish NCN grant DEC-2011/03/B/ST9/03459.
We made use of data and software provided by the Fermi Science Support Center, managed by the HEASARC at the Goddard Space Flight Center, and Swift/BAT transient monitor results provided by the Swift/BAT team.
\end{acknowledgments}

\bigskip 
{\bf References}

\begin{enumerate}

\item
Acero F., et al.\ 2015, ApJSS, submitted

\item
Ackermann M., et al.\ 2012, ApJ, 747, 104

\item
Krimm H. A., et al.\ 2013, ApJSS 209, 14. 

\item
Lubi\'nski P., Zdziarski A.~A., Walter R., Paltani S., Beckmann V., Soldi S., Ferrigno C., Courvoisier T.~J.-L.\
2010, MNRAS, 408, 1851

\item
Nied{\'z}wiecki A., St{\c e}pnik A., Xie F.-G.\ 2015, ApJ, 799, 217 

\item
Nied{\'z}wiecki A., Xie F.-G., St{\c e}pnik A.\ 2013, MNRAS, 432, 1576

\item
Nolan P.~L.,  et al.\ 2012, ApJSS, 199, 31 

\item
Oka K., Manmoto T.\ 2003, MNRAS, 340, 543 

\item
Onken C.~A., et al.\ 2014, ApJ, 791, 37 

\item
Teng S.~H., Mushotzky 
R.~F., Sambruna R.~M., Davis D.~S., 
Reynolds C.~S.\ 2011, ApJ, 742, 66 

\item
Wojaczy\'nski R., et al.\ 2015, in preparation 

\item
Yuan F., \& Narayan R.\ 2014, ARAA, 52, 529 

\end{enumerate}

\end{document}